\documentclass[twocolumn, pra, superscriptaddress]{revtex4-1}
\usepackage{upgreek}
\usepackage{tikz}
\usepackage{dcolumn}
\usepackage{color,tcolorbox}
\usepackage{amsmath,mathrsfs,amssymb,esint}
\usepackage{dcolumn}
\usepackage{bm}
\usepackage{blindtext}
\usepackage{natbib}
\usepackage{graphicx}
\usepackage{rotating}
\usepackage{titlesec}
\usepackage{multirow}
\usepackage{mathtools}
\usepackage[title]{appendix}
\usepackage[pdfborder={0 0 0},colorlinks=true,linkcolor=blue,urlcolor=blue,citecolor=blue]{hyperref}
\usepackage{enumitem}

\DeclareMathOperator{\sech}{sech}
\colorlet{orange1}{green!10!orange!90!}
\definecolor{ForestGreen}{rgb}{0,0.5,0}

\usepackage[T1]{fontenc}
\usepackage{xr}
\usepackage{xcite}
\externalcitedocument{liu_et_al_prr_resub_appendix}
\begin{document}
\title{Phase Diagram of Solitons in the Polar Phase of a Spin-1 Bose-Einstein Condensate}
\author{I-Kang Liu}
\email{i-kang.liu1@newcastle.ac.uk}
\affiliation{
School of Mathematics, Statistics and Physics, Newcastle University, Herschel Building, Newcastle upon Tyne, NE1 7RU, United Kingdom}
\affiliation{Department of Physics and Graduate Institute of Photonics, National Changhua University of Education, Changhua 50058, Taiwan}

\author{Shih-Chuan Gou}
\affiliation{Department of Physics and Graduate Institute of Photonics, National Changhua University of Education, Changhua 50058, Taiwan}
\author{Hiromitsu Takeuchi}
\email{hirotake@sci.osaka-cu.ac.jp}
\affiliation{Department of Physics and Nambu Yoichiro Institute of Theoretical and Experimental Physics (NITEP), Osaka City University, Osaka 558-8585, Japan}
\date{\today}
\begin{abstract}
We theoretically study the structure of a stationary soliton in the polar phase of spin-1 Bose--Einstein condensate in the presence of quadratic Zeeman effect at zero temperature. The phase diagram of such solitons is mapped out by finding the states of minimal soliton energy in the defining range of polar phase. The states are assorted into normal, anti-ferromagnetic, broken-axisymmetry, and ferromagnetic phases according to the particle and spin densities in the core. The order of phase transitions between different solitons and the critical behaviour of relevant continuous transitions are proved within the mean-field theory.
\end{abstract}
\maketitle
\section{Introduction}
Generation of topological defects caused by spontaneous symmetry breaking (SSB) is a universal phenomenon constantly addressed in cosmology, high energy and condensed matter physics \cite{2000Bunkov,Kibble76,Zurek96}.
Depending on the type of SSB transition and the degrees of freedom of order parameters topological defects are created in various forms, such as vortices/strings~\cite{2013Lamporesi,Liu2018}, domain walls/solitons~\cite{2010Damski}, hedgehogs/monopoles, or their composites.
In the past decades, the multi-component or spinful superfluids of liquid $^3$He \cite{2003Volovik,2013Vollhardt,Volovik2020} and gaseous Bose-Einstein condensates (BECs) \cite{Dalfovo1999,Pethick2008,Kawaguchi2012}, which induce SSB in numerous ways because of their multiple degrees of freedom of order parameters, have served as a testing ground for the creation of novel topological defects. 
In these multi-component systems, the core of a defect is not necessarily singular.
A simple example is the so-called {\it coreless} vortex in which the vortex core of a superfluid component is occupied by other components~\cite{2003Leahrdt}, in contrast to the vortex in a single-component superfluid, whose core is singular due to the nonexistence of superfluid order. Another remarkable example is the core structure of vortices in superfluid $^3$He-B~\cite{1995Parts}, where vortices undergo a phase transition by changing the core structure depending on the pressure and temperature.

Recently, Kang {\it{et~al.}}~\cite{Kang2018} observed the wall-vortex composite defects spontaneously generated in a quasi two-dimensional (2D) spinor BEC of sodium quenched from the antiferromagnetic phase to the polar~(P) phase~\cite{AFandP}.
In the early stage of evolution, the spontaneous breaking of discrete symmetry causes the formation of domain walls as dark solitons in the $|m=0\rangle$ Zeeman component.
In contrast to the dark soliton in a scalar BEC~\cite{Anderson2001}, whose lifetime is generally short, the wall structure exists for a long time with its core occupied by the $|m=\pm 1\rangle$ Zeeman components, forming a composite defect of wall and half-quantum vortices~\cite{Kang2018}.
This suggests that, owing to its solitary nature, soliton not only acquires a role in one-dimensional (1D) systems~\cite{Burger1999,Busch2001,Frantzeskakis2010,2010Damski,2011Witkowska} but can also serve as the building block of some exotic composite defects in the multi-component systems in a higher dimension.
Despite that the experiment in Ref.~\cite{Kang2018} has revealed certain dynamical features of the solitons in a spinor BEC, a theoretical investigation of such solitons in the presence of quadratic Zeeman effect is lacking~\cite{2005Li,2006Uchiyama,2008Nistazakis,2018Bersano,Meng2019}.

\begin{figure}[t!]
\centering
\includegraphics[width=1\linewidth]{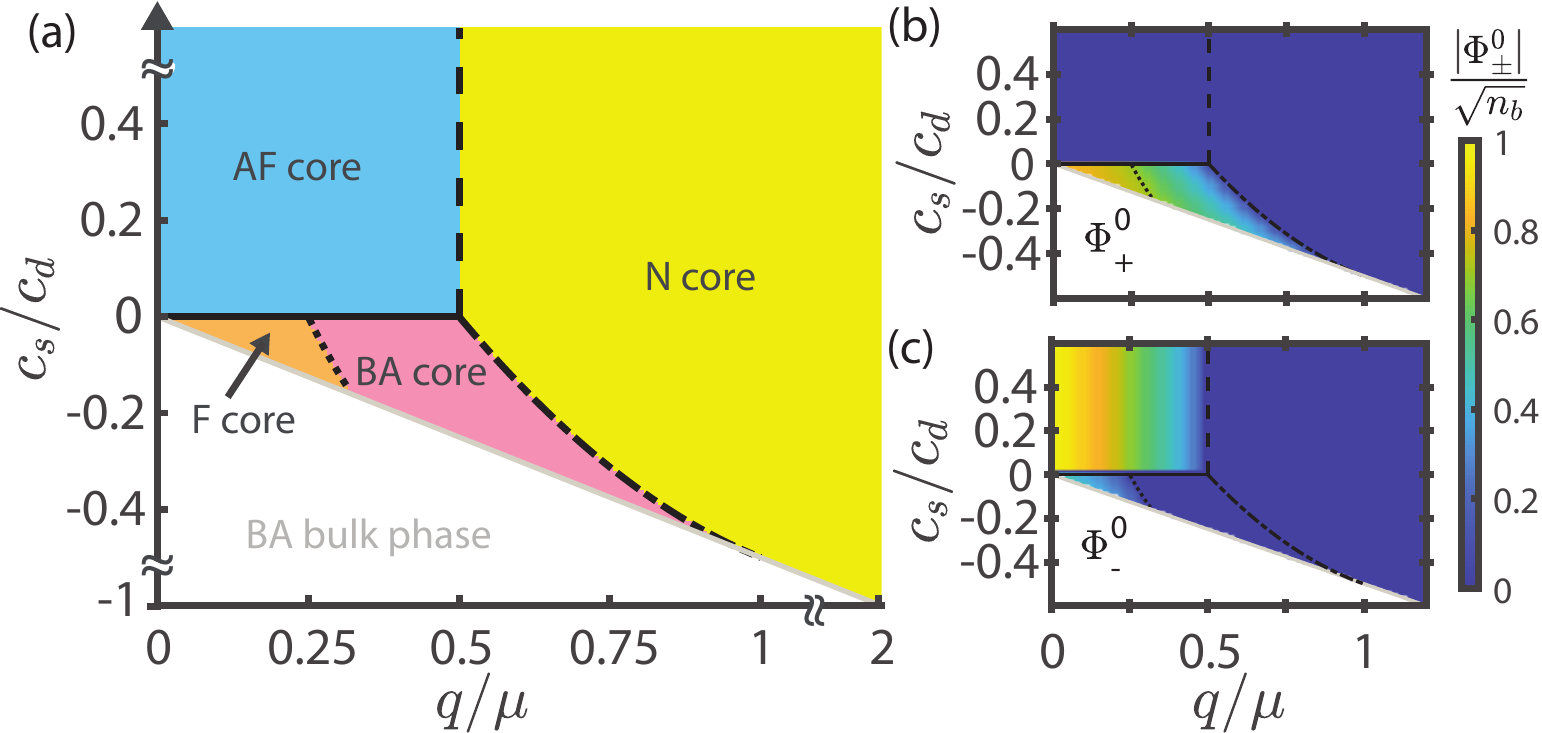}
\caption{(a) The typical phase diagram of stationary soliton states in the defining range of polar phase ($p=0$).
(b) and (c) Variations of the order parameters $\Phi_\pm^0=\Phi_\pm(x=0)$. The phase boundary drawn in black solid line [AF-F and AF-BA in (a)] indicates a phase transition of first order, across which the order parameter changes discontinuously. On the other hand, the phase boundary drawn in broken line [AF-N, F-BA and BA-N in (a)] indicates a phase transition of second order, across which the order parameter changes continuously.
}
\label{fig:1}
\end{figure}

In this paper, we investigate the core structure of solitons in the P phase of spin-1 BECs, using the Gross-Pitaevskii formalism.
Our findings unfold the less-explored aspects of soilton physics in the multi-component superfluid systems.
The remaining part of this paper is organized as follows.
In Sec.~\ref{sec:formulae}, the formulae for calculating soliton energy are presented.
In Sec.~\ref{sec:results}, the phase diagram, featuring a variety of stationary states distinguished by the particle and spin densities, is obtained and illustrated in Fig.~\ref{fig:1}~(a).
Upon identifying the phase boundaries, we determine the order of phase transition between two bordering phases and further analyze the critical behaviour of those continuous phase transitions by applying a Ginzburg-Landau-like theory.
Finally, some concluding remarks are given in Sec.~\ref{sec:remarks}.

\section{Gross-Pitaevskii theory of spin-1 Bose gases}\label{sec:formulae}
It is well received that the mean-field theory applies particularly successfully to a weakly interacting Bose gas.
In the context of BEC, the Gross-Pitaevskii~(GP) theory renders a classical description of Bose-condensed systems by ignoring all fluctuations.
In most cases, the GP theory describes the static as well as dynamic properties of a weakly interacting BEC in excellent agreement with experimental oberavations~\cite{Becker2008,Weller2008,*Frantzeskakis2010,*Theocharis2010,*Kevrekidis2016,Farolfi2019,Chai2019}, provided that the temperature is sufficiently low~\cite{Dalfovo1999,Lieb2000,Pethick2008,mean_field}.
In this section, we give a brief account for the GP theory of spin-1 BEC.
To begin with, we note that a spin-1 BEC is described by a vectorial macroscopic wave function, $\vec{\Psi} (\mathbf{r},t)=(\Psi_1,\Psi_0,\Psi_{-1})^T$~\cite{Kawaguchi2012}.
A stationary solution of $\vec{\Psi}$ corresponds to the local minimum of the thermodynamic energy functional $E[\vec{\Psi}]=\int d\mathbf{r}\mathcal{E}[\vec{\Psi}]$ with the energy density given by
\begin{equation}\begin{array}{rl}
\mathcal{E}[\vec{\Psi}]=&\displaystyle\sum_{m=-1}^1\Psi_m^\ast\left[-\frac{\hbar^2\nabla^2}{2M}-pm+qm^2\right]\Psi_m
\\
&\displaystyle\quad-\mu n+\left[\frac{c_d}{2}n^2+\frac{c_s}{2}\left|\mathbf{f}\right|^2\right]
\end{array}
\label{eq:therm_E}
\end{equation}
where $M$ is the atomic mass, $p$ and $q$ the linear and quadratic Zeeman coefficients respectively, $\mu>0$ the chemical potential, $n=\sum_{m=-1}^1|\Psi_m|^2$ the particle density, and $\mathbf{f}=(f_x,f_y,f_z)^T$ is the spin density vector with $f_x=\sqrt{2}\textrm{Re}[(\Psi_1+\Psi_{-1})\Psi_0^\ast]$, $f_y=\sqrt{2}\textrm{Im}[(\Psi_1-\Psi_{-1})\Psi_0^\ast]$ and $f_z=|\Psi_1|^2-|\Psi_{-1}|^2$. The constants $c_d$ and $c_s$ denote separately the strengths of density-density and spin-spin interactions mediated by $s$-wave collisions. In this paper, we assume $p=0$ ~\cite{Kang2018}, and thus the P phase is now delimited by $2\geq q/\mu>0$ and $c_s/c_d>-q/2\mu>-1$~\cite{Kawaguchi2012}.
Without loss of generality, the ground state of spin-1 BEC is represented by $\vec{\Psi}_P=(0,\sqrt{n_b},0)$, where $n_b$ is the bulk density. Note that the arbitrariness of the global phase ensures that $\vec{\Psi}_P$ can be real-valued.
The chemical potential $\mu=c_d n_b$ and the density healing length $\xi=\hbar/\sqrt{M\mu}$ characterize the energy and length scales of P phase.

Consider a flat domain wall normal to the $x$-axis in a uniform spinor condensate. The translational invariance in the $y$- and $z$-directions renders the problem into a 1D case. Denoting the wave function as $\vec{\Psi}_{\textrm{soliton}}(x)=(\psi_1,\psi_0,\psi_{-1})^T$, and assuming that the soliton is centered at $x=0$, then in analogy to the dark soliton in the scalar BEC, $\psi_0$ obeys the boundary condition $\psi_0(x=\pm \infty) =\pm\sqrt{n_b}$, imposing a $\pi$-phase jump across the soliton core. We also assume that $\vec{\Psi}_{\textrm{soliton}}$ produces no currents, i.e., $\mathbf{j}_m=\left|\psi_m\right|^2\nabla\textrm{arg}(\psi_m)=0$ for the stationary solution~\cite{stationaryHere}. Although $\psi_0$ vanishes at the core center, the $\psi_{\pm1}$ components can readily occupy the core region, forming the so-called bright-dark-bright soliton~\cite{Salasnich2006,2008Nistazakis,Becker2008}.
Note that ${E}[\vec{\Psi}]$ is invariant under a rotational transformation of $\mathbf{f}$ about the $z$-axis. We thus choose $(f_x,f_y)=(f_\perp,0)$ for simplicity. Consequently, the fields $\psi_m(x)$ can be real-valued.

The soliton solution with lowest energy is determined by minimizing the soliton energy,
\begin{equation}
\alpha[\vec{\Psi}_{\textrm{soliton}}]=\int dx\left\{\mathcal{E}[\vec{\Psi}_{\textrm{soliton}}]-\mathcal{E}[\vec{\Psi}_P]\right\}.
\label{eq:soliton_energy}
\end{equation}
This quantity corresponds to the excess energy per unit area of the soliton,
which plays the role of the tension coefficient of domain wall \cite{Landau1980}.
Requiring that $\vec{\Psi}_{\text{soliton}}$ minimizes the soliton energy, leads to the following coupled GP equations,
\begin{equation}\begin{array}{rl}
\mu\psi_0=&\displaystyle\left[-\frac{\hbar^2\partial_x^2}{2M}+c_dn\right]\psi_0+\frac{c_s}{\sqrt{2}}f_x(\psi_1+\psi_{-1})
\\\\
\mu\psi_{\pm1}=&\displaystyle\left[-\frac{\hbar^2\partial_x^2}{2M}+q+c_dn\pm c_sf_z\right]\psi_{\pm1}+\frac{c_s}{\sqrt{2}}f_x\psi_0.
\end{array}
\label{eq:GPE}
\end{equation}
Using imaginary-time propagation method, the desired solutions are obtained by numerically solving Eq.~(\ref{eq:GPE}) subjected to the Neumann boundary condition $\partial_{x}\psi_m(x=\pm L/2)=0$, where the system size $L$ is sufficiently large compared to the width of the core~\cite{NumericalDetails1}.

\begin{figure}[t!]
\centering
\includegraphics[width=1\linewidth]{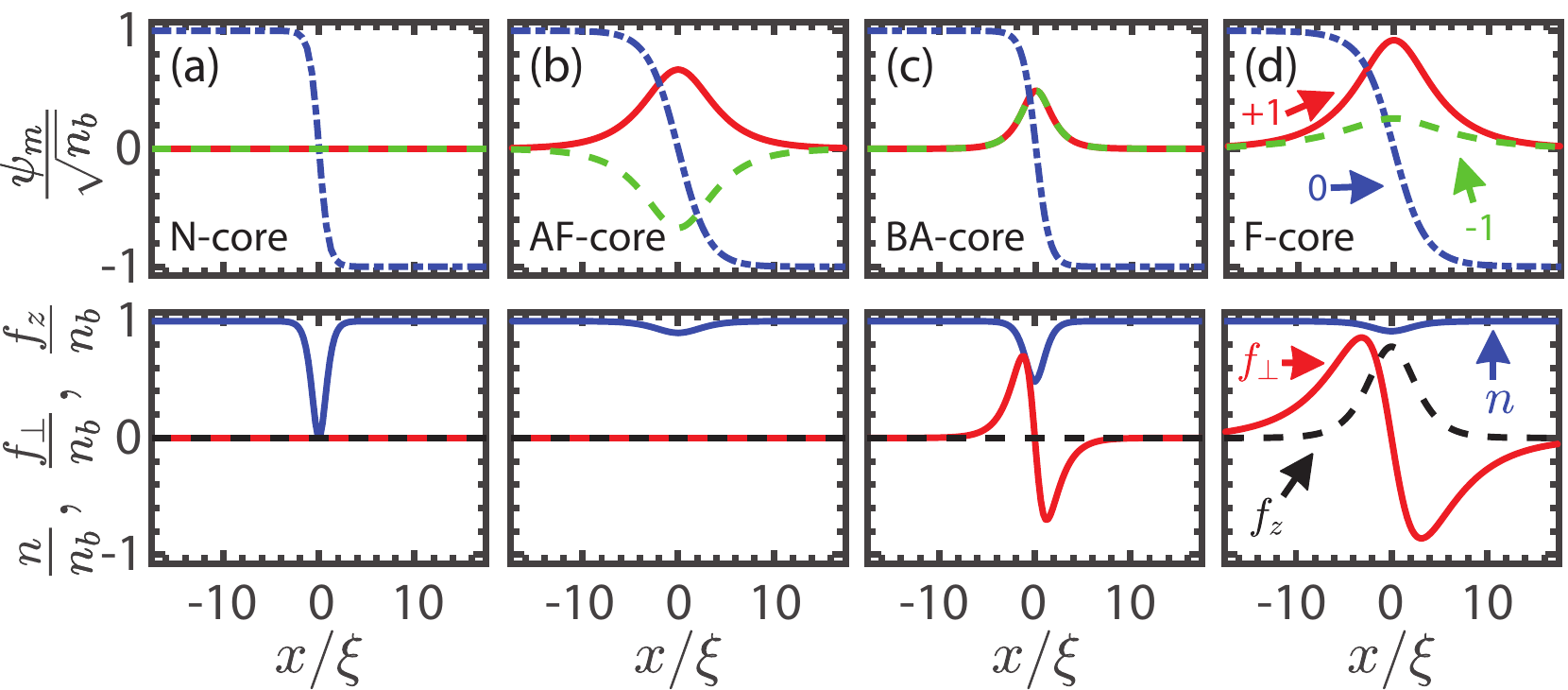}
\caption{The typical core profiles of the soliton solutions with the lowest soliton energy.
The profiles of the macroscopic wave functions $\psi_m$ are plotted in the top panels for (a) N-core with $(q/\mu,c_s/c_d)=(0.5125, 0.0125)$, (b) AF-core with $(q/\mu,c_s/c_d)=(0.05,0.0125)$, (c) BA-core with $(q/\mu,c_s/c_d)=(0.2625,-0.0125)$ and (d) F-core with $(q/\mu,c_s/c_d)=(0.05, -0.0125)$.
The bottom panels show the corresponding particle density $n$, transverse spin density $f_\bot$, and axial spin density $f_z$.
}
\label{fig:2}
\end{figure}
\section{Results and Discussions}\label{sec:results}
\subsection{Phase diagram}
Figure~\ref{fig:1}~(a) shows the phase diagram of soliton obtained numerically by minimizing the soliton energy.
Here, by rescaling energy and length by $\mu$ and $\xi$, the number of independent parameters reduces to two, namely, $q/\mu$ and $c_s/c_d$. 
We conclude that the core structure of these soliton states can be distinguished into four types: normal (N), antiferromagnetic (AF), broken-axisymmetry (BA) and ferromagnetic (F), as shown in Fig.~\ref{fig:2}.
The N-core is prescribed by $\psi_{\pm 1}(x)=0$, or equivalently, $\psi_0$ is precisely the dark-soliton solution in the usual scalar BECs~\cite{Pethick2008}, such that $\vec{\Psi}_{\text{soliton}}$ of this kind features the "normal (N) state" at its core center as shown in Fig.~\ref{fig:2}~(a). 
The AF-core illustrated in Fig.~\ref{fig:2}~(b) is just the case observed in Ref.~\cite{Kang2018}, which is occupied by the "AF state" with $\psi_{1}=-\psi_{-1}$, i.e., $\psi_{1}$ and $\psi_{-1}$ are out of phase. Correspondingly, the BA-core is occupied by the "BA state" with $\psi_{1}=\psi_{-1}$, i.e., $\psi_{1}$ and $\psi_{-1}$ are in phase. We notice that, in contrast to the N- and AF-cores whose spin density vectors $\mathbf{f}$ vanish everywhere, the spins of the BA-core are all aligned in the transverse direction with a node at the core center, in other words, $f_z=0$, as shown in Fig.~\ref{fig:2}~(c). Finally, the F-core illustrated in Fig.~\ref{fig:2}~(d) can be considered as a generalization of BA-core, which is occupied by the "F state" with $f_z\neq 0$ in the core. The F-core features nonzero spin distribution in both transverse and axial directions.

The phase diagram can be understood qualitatively by considering two length scales associated with the thickness of solitons.
For simplicity, let us first consider the case of magnetization-free AF-N transition for $c_s>0$.
Then, the relevant length scales in this case are related to the chemical potential and quadratic Zeeman term by $\xi=\hbar/\sqrt{M\mu}$ and $\xi_q=\hbar/\sqrt{Mq}$, respectively~\cite{Kang2018}. For the AF-core, we have $\xi_q>\xi$ in the regime of small $q/\mu$, and the $\psi_{\pm1}$ components are largely accommodated in the core region.
With increasing $q/\mu$, $\xi_q$ becomes comparably smaller than $\xi$, and both $\psi_{\pm1}$ components in the AF-core diminish until they totally vanish before transitioning to the N-core.
As a result, we conclude that core size$\sim \max(\xi,\xi_q)$.
For the ferromagnetic interaction ($c_s<0$), the above argument on the length scales becomes intricate as the nonzero $\mathbf{f}$ of the F- and BA-cores would introduce additional length scale related to the spin-spin interaction.

To probe into the essence of the phase transition between different types of soliton, it is necessary to appeal to other more rigorous approaches, and to this end, we introduce the real scalar fields which are defined as
\begin{equation}
\left(\Phi_0, \Phi_{\pm}\right)=\left(\psi_0,\frac{\psi_{+1}\pm\psi_{-1}}{\sqrt{2}}\right).
\label{eq:Phi}
\end{equation}
Accordingly, the particle and spin densities are respectively expressed as $n=\Phi_+^2+\Phi_-^2+\Phi_0^2$ and $(f_\bot,f_z)=(2\Phi_+\Phi_0,2\Phi_+\Phi_-)$. The states in the phase diagram can be distinguished by (1) N-core: $\Phi_{\pm}=0$; (2) AF-core: $\Phi_-\neq 0$ and $\Phi_+=0$; (3) BA-core: $\Phi_+\neq 0$ and $\Phi_-=0$; (4) F-core: $\Phi_+\neq\Phi_-\neq0$. The validity of $\Phi_\pm$ is shown in Fig.~\ref{fig:1}~(b) and (c), where $\Phi_\pm^0=\Phi_\pm(x=0)$ are evaluated in the defining range of P phase and are demonstrated in false colors, with the phase boundaries in Fig.~\ref{fig:1}~(a) faithfully matched. 

We note that there is a jump of $\Phi^0_\pm$ across the horizontal axis, $c_s=0$, implying that both of the AF-F and AF-BA phase transitions are discontinuous or first order. On the other hand, $\Phi^0_\pm$ varies continuously along the horizontal-axis: the AF-N, F-BA and BA-N phase transitions are continuous or second order.
We investigate the order of phase transition by computing the derivative of the soliton energy with respect to $q/\mu$ and $c_s/c_d$, and the conclusion is consistent with the above observation~\cite{NumericalDetails1}.


\subsection{Perturbative approach}
In the representatoin of $\Phi_\pm$, the coupled equations of $\psi_{\pm1}$ are turned into
\begin{equation}
-q\Phi_\pm=\displaystyle\left[-\frac{\hbar^2\partial_x^2}{2M}+c_dn+2c_s\Phi_\mp^2+(1\pm1)c_s\psi_0^2-\mu\right]\Phi_\pm,
\label{eq:Phi_GPE}
\end{equation}
In the close vicinity to the left of the N-core region in the phase diagram, $\Phi_\pm$ are perturbatively small in magnitude and we can assume $\psi_0=\sqrt{n_b}\tanh (x/\xi)$. Similar to the case of coreless vortices in the segregated binary condensates~\cite{Hayashi2013}, we neglect the cubic terms in the right-hand-side of Eq.~(\ref{eq:Phi_GPE}), 
and the core-occupying components can be approached by the exact solution of P\"{o}schl-Teller equation~\cite{Diaz1999,*PTequation},%
\begin{equation}
E_\pm\Phi_\pm(x) =\left[ -\frac{\hbar^2\partial_x^2}{2M}+V_\pm(x) \right]\Phi_\pm(x),
\label{eq:linearPhi}
\end{equation}
where $E_\pm=-q-(1\pm1)({c_{s}}/{c_{d}})\mu $ are the eigenvalues determined by the potentials $V_\pm(x)=-U_\pm\sech^2(x/\xi)$ respectively, with $U_\pm=\left[ 1+(1\pm1)({c_{\rm s}}/c_{\rm d})\right]\mu$.

The above perturbative treatment suggests that $\Phi_\pm$ respectively correspond to the bound-state solutions of lowest energy eigenvalues $\epsilon_{\pm,0}$ associated with $V_\pm$ in Eq.~(\ref{eq:linearPhi}), i.e., 
\begin{equation}
\epsilon_{\pm,0}=-\frac{\mu}{8}\left[\sqrt{1+\frac{8M\xi^2U_{\pm}}{ \hbar ^{2}}}-1\right]^{2},
\end{equation}
where $\sqrt{1+8M\xi^2U_\pm/\hbar^2}>1$ with $c_s/c_d>-1/2$ for $V_+$~\cite{PTtheory}.
The phase boundaries of the AF-N and BA-N transitions are determined by $E_\pm=\epsilon_{\pm,0}$ and thus we obtain equations for the phase boundaries,
\begin{equation}\label{eq:BoundaryEq}
\begin{cases}
\dfrac{q}{\mu } =\dfrac{1}{2}, & \textrm{ for } \Phi_- \textrm{ (AF-N boundary)}\\
\dfrac{c_{s}}{c_{d}} =1-\dfrac{5}{2}\dfrac{q}{\mu } +\dfrac{q^{2}}{\mu ^{2}}, & \textrm{ for }\Phi_+ \textrm{ (BA-N boundary})
\end{cases}.
\end{equation}
Remarkably, the above equations coincide closely with our numerical results in Fig.~\ref{fig:1}. In particular, the condition $c_s/c_d>-1/2$ is coincidentally satisfied since the BA-N boundary terminates at $(q/\mu,c_s/c_d)=(1,-1/2)$ on the boundary of the bulk BA phase.


\subsection{Flat-core limit}
To gain an insight into the small $q$ regime, where $\psi_{\pm1}$ are barely suppressed in the core in the phase diagram, we introduce a variational analysis assuming the flat-core limit $n=\textrm{const}.$
In the limit $q\rightarrow0^+$,
the soliton solution can be well approximated by the bright-dark-bright soliton solution in multicomponent condensates~\cite{Salasnich2006,Busch2001,2008Nistazakis,Becker2008}%
\begin{equation}\label{eq:psi_v}
\vec{\Psi}_v =\sqrt{n_b}\left(\cos \phi _v {\rm sech}\frac{x}{\xi _v} ,\tanh\frac{x}{\xi _v} ,\sin \phi _v{\rm sech}\frac{x}{\xi _v}\right)^T.
\end{equation}
Here $\phi_v$ ($0\leq\phi_v\leq\pi$) and $\xi_v$ are variational parameters, and the total density is normalized to the constant bulk density $n_b$. Substituting $\vec{\Psi}_v$ into Eq.~(\ref{eq:soliton_energy}), the soliton energy is explicitly expressed by 
\begin{equation}
\alpha[\vec{\Psi}_v]=\mu n_b\xi\left\{\frac{\xi}{\left|\xi_v\right|}+2\frac{|\xi_v|}{\xi}\left[
\frac{q}{\mu}+\frac{c_s}{3c_d}h(\phi_v)
\right]
\right\},
\label{eq:variational_alpha}
\end{equation}
where $h(\phi_v)=1+\sin(2\phi_v)+\cos^2(2\phi_v)$.
Minimizing $\alpha[\vec{\Psi}_v]$ in Eq.~(\ref{eq:variational_alpha}) with respect to $\xi_v$ and $\phi_v$, we obtain the following energetically favorable states:
(i) AF-core with $(\phi_v,\xi_v)=(-\pi/4,\pm\xi\sqrt{\mu/2q})$ for $c_s>0$; (ii) F-core with $(\phi_v,\xi_v)=(\pi/4\pm\pi/6,\pm \xi/\sqrt{(2q/\mu+3c_s/2c_d}))$ for $c_s<0$; (iii) BA-core with$ (\phi_v,\xi_v)=(\pi/4,\pm\xi/\sqrt{(2q/\mu+4c_s/3c_d)})$, which is a local minimum (saddle point) for $c_s>0$ ($c_s<0$).
Note that these states are all doubly degenerate~\cite{flat_core_var_para}.

\begin{figure}[t!]
\centering
\includegraphics[width=1\linewidth]{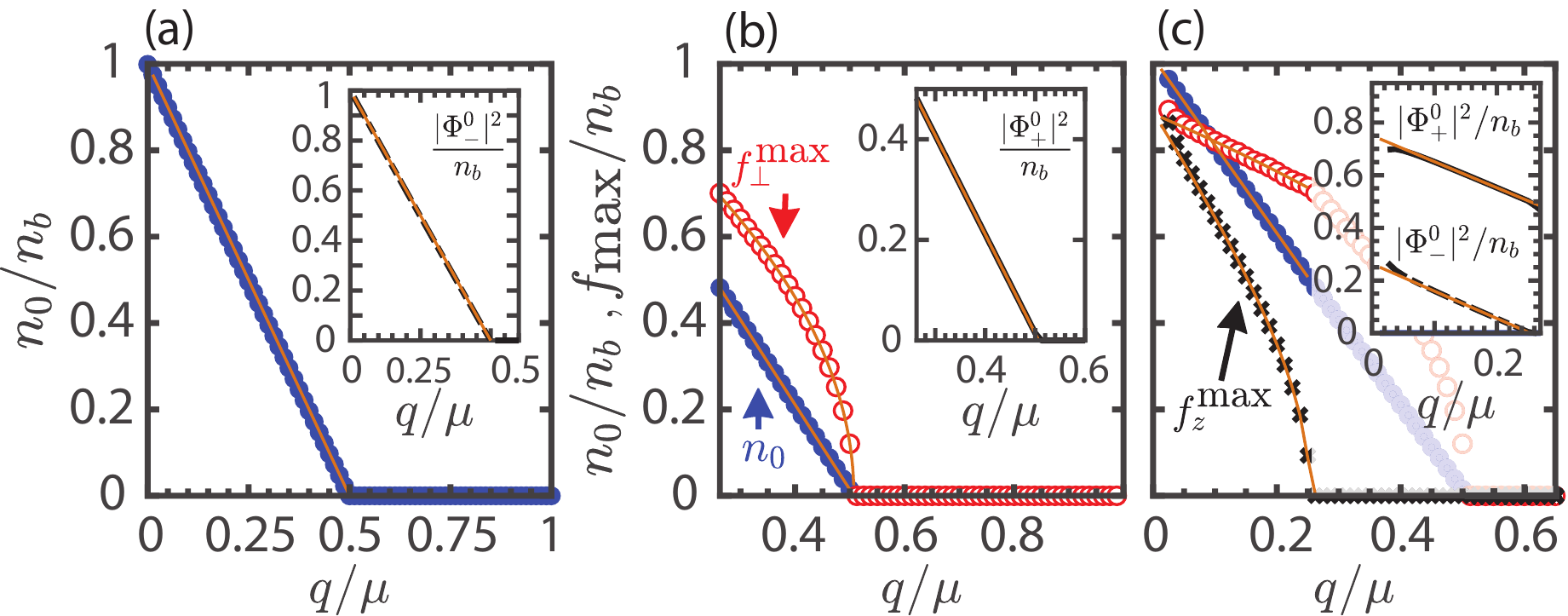}
\caption{Numerical simulations for the critical behavior of $n_0$ (blue circle), $f_{\perp}^\text{max}$ (red circle), $f_{z}^\text{max}$ (black cross) and their best fits (orange solid lines). (a) AF-N transition, fitting parameters: $c_s>0$, $q_c=0.5\mu$ and $\beta=2n_b/\mu$; (b) BA-N transition, fitting parameters:$c_s=-0.0125c_d$, $q_c=0.51\mu$ and $\beta=1.97n_b/\mu$; (c) F-BA transition, fitting parameters: $c_s=-0.0125c_d$, $\beta=1.04n_b/\mu$, $q_c=0.25\mu$, $B_0=0.5$ and $B_1=0.99$. }
\label{fig:3}
\end{figure}

The above variational analysis theoretically justifies the small-$q$ regime of the phase diagram, and, in particular, accounts for the existence of the F-core soliton.
The analysis also describes $q/\mu-$ and $c_s/c_d$-dependence of the soliton thickness $\xi_v$;
the thickness could depend on the spin-spin interaction through $c_s/c_d$ in addition to the quadratic Zeeman effect for $c_s<0$ as mentioned before.

\subsection{Critical behaviour of the observables}

To this far, we see that the soliton can undergo a continuous phase transition from AF- to N-core, or successive continuous phase transitions from F- to BA- and then to N-core. Figure~\ref{fig:1}~(b) and (c) reveal that these phase transitions are also accompanied by the vanishing of $\Phi_+^0$ or $\Phi_-^0$ at a critical point $q=q_c$, which suggests that $\Phi_{\pm}^0$ can serve as the effective order parameters for the continuous transition. In what follows, we employ a mean-field analysis similar to the Ginzburg-Landau theory~\cite{Landau1980} to probe the critical behaviour of the continuous phase transition between different core structures.

We expand the soliton energy in terms of the effective order parameter $\varphi$ near the critical point ($q\leq q_c$) as $\alpha=\alpha_0-\alpha_{1}(q-q_c)\left|\varphi\right|^2-\alpha_{2}\left|\varphi\right|^4$, with $\alpha_i \ge 0$ ($i=0, 1, 2$). Here $\varphi=\Phi_-^0$ for the AF-N and F-BA transition, and $\varphi=\Phi_+^0$ for the BA-N transition. Minimization of $\alpha_c$ with respect to $\varphi$ yields $\left|\varphi\right|^2=\beta\left|q-q_c\right|$ with $\beta=\alpha_{1}/\alpha_2$, which manifests itself as a linear ramping down of the central particle density, $n_0=n(x=0)=\left|\varphi\right|^2=\left|\Phi_\pm^0\right|^2\propto\left|q-q_c\right|$, in the critical regions of the AF-N and BA-N transitions, respectively, as demonstrated in Fig.~\ref{fig:3}~(a)-(b). The parameter $\beta$ can be determined by linear fitting of $n_0$. Likewise, the maximal spin density around the core demonstrates a power law in the critical region. As is derived in the Sec.~\ref{sec:critical_behaviour} of supplemental material for the BA-N transition, $f^\text{max}_{\bot}=\textrm{max}[f_{\bot}(x)]\approx\Phi_+^0\propto\sqrt{\left|q-q_c\right|}$. However, such power-law behavior of spin densities does not arise in the AF-N transition, as $\mathbf{f}$ vanishes for both AF- and N-cores.

The analysis for the F-BA transition is delicate. As shown in the inset of Fig.~\ref{fig:3}~(c), both $\left|\Phi_\pm^0\right|^2$ vary virtually linearly in the critical region. We notice that while $\left|\Phi_-^0\right|^2$ vanishes at $q_c$, $\left|\Phi_+^0\right|^2$ decreases to a nonzero value at $q_c$. Thus we assume $\left|\Phi_+^0\right|^2=B_0+B_1\left|q-q_c\right|$ for $q\leq q_c$, where $B_0, B_1>0$, and the central density is given by $n_0=B_0+(\beta+B_1)\left|q-q_c\right|$. Here the coefficients $\beta$, $B_0$ and $B_1$ can be determined from the linear fitting of $n_0$. Furthermore, it is straightforward to show that $f_\bot^\text{max}\propto\sqrt{B_0+B_1\left|q-q_c\right|}$ and $f_{z}^\text{max}\propto\sqrt{\beta\left|q-q_c\right|\left(B_0+B_{1}\left|q-q_c\right|\right)}$, which nicely fit to the numerical simulations of $f_\bot^\text{max}$ and $f_{z}^\text{max}$, as demonstrated in Fig.~\ref{fig:3}~(c). We notice that, in the close vicinity of $q_c$, the power-law behavior $f_{z}^\text{max}\propto\sqrt{\left|q-q_c\right|}$ is resumed.

\section{Concluding Remarks}\label{sec:remarks}
We have theoretically inquired into the core structure of stationary solitons in the P phase of spin-1 BECs. In the presence of quadratic Zeeman effect, four different types of structure are identified according to the particle, spin densities in the core.
Our theoretical analyses support well the numerically obtained phase diagram of solitons.
The critical behaviour of continuous phase transition between different types of solitons is predicted by introducing the Ginzburg-Landau-like mean-field theory.

As the AF-core soliton was already created in the spinor condensate of sodium (with $c_s>0$)~\cite{Kang2018}, the F- and BA-core solitons can in principle be generated in a spinor condensate of rubidium/lithium (with $c_s<0$) via quantum quenches~\cite{Kang2018,Farolfi2019,Chai2019,Prfer2018,*JimnezGarca2019,*Huh2020} or phase imprinting method~\cite{Chai2019,Lannig2020}, which are within the reach of the state-of-the-art techniques. 
Furthermore, by tuning the quadratic Zeeman energy via changing the magnitude of magnetic field, the transition between different types of soliton states can be probed experimentally. Our study thus offers possibilities to explore different aspects of soliton physics, e.g., magnetic phase transition and spontaneous symmetry breaking of soliton cores. One of the follow-up studies would be the phase diagram of solitons in other experimentally accessible phases of spinor BECs. Dynamical properties of solitons, such as the instability or collisions between solitons, are also important issues to be looked into while exploring the non-equilibrium dynamics in multi-component superfluids.


{\bf{Acknowledgement}} I.-K. Liu and S.-C. Gou were financially supported by MOST 106-2112-M-018-005-MY3 (Taiwan).
H. Takeuchi has been supported by JSPS KAKENHI Grant Numbers JP17K05549, JP18KK0391, JP20H01842, and in part by the OCU "Think globally, act locally" Research Grant for Young Scientists 2019 and 2020 through the hometown donation fund of Osaka City.
We appreciate the fruitful discussions with Yong-il Shin, Yu-Ju Lin and Thomas Bland.

\appendix
\begin{appendices}
\section{: Numerical analysis}\label{sec:numerical_process}
For a homogeneous spin-1 condensate in the P phase, we choose the chemical potential $\mu=c_dn_b$ ($n_b$ is a constant density)~\cite{Kawaguchi2012} as the characteristic energy. Accordingly, the characteristic length and time scales are defined as $\xi=\hbar/\sqrt{M\mu}$ (the healing length) and $\tau=\hbar/\mu$, respectively. By rescaling the energy functional,  Eq.~(\ref{eq:therm_E}), with respect to these variables, one obtains the dimensionless energy functional in the absence of linear Zeeman term,
\begin{equation}\begin{array}{rl}
E'[\vec{\Psi}']=&\displaystyle\sum_{m=-1}^1\int d\mathbf{r}'\Psi_{m}'^\ast\left[
-\frac{\nabla'^2}{2}+q'm^2-1\right]\Psi_{m}'
\\\\
&\displaystyle+\frac{1}{2}\int d\mathbf{r}'\left[n'^2+c_s'|\mathbf{f}'|^2\right],
\end{array}
\label{eq:1D_dimensionless_energy}
\end{equation}
where ${\bf r'}={\bf r}/\xi$, $E'={E}/{(\mu n_b \xi^3)}$, $q'=q/\mu$, $c_s'=c_s/c_d$. The scaled wave function is related to the original one by $\Psi'_m=\Psi_m/\sqrt{n_b}$.  Accordingly, the dimensionless total particle density is given by $n'=\sum_{m=-1}^1|\Psi_m'|^2$, and the scaled spin-density vector are given by is $\mathbf{f}'=f'_x\mathbf{e}_x+f_y'\mathbf{e}_y+f_z'\mathbf{e}_z=\mathbf{f}_\perp'+f_z'\mathbf{e}_z$ where $f_{x}'=\sqrt{2}\textrm{Re}[\Psi_0'^\ast(\Psi_1'+\Psi'_{-1})]$, $f_y=\sqrt{2}\textrm{Im}[\Psi_0'^\ast(\Psi_1'-\Psi_{-1}')]$ and $f_z'=|\Psi_1'|^2-|\Psi_{-1}'|^2$.
 
Now we consider a soliton solution by chossing $x$-axis as the coordinate normal to the wall with the wave function, $\vec{\Psi}_{\rm solition}'=(\psi_1'(x'),\psi_0'(x'),\psi_{-1}'(x'))^T$. Consequently, the dimensionless soliton energy is
\begin{equation}
\alpha'=\frac{\alpha}{\mu n_b\xi}=E_{\textrm{soliton}}'[\vec{\Psi}_{\rm soliton}']-E_{\textrm{bulk}}'.
\end{equation}
where $E'_{\textrm{bulk}}=L/2\xi=L'/2$. The corresponding dimensionless GP equations are derived by using Hartree variational principle, $\delta \alpha'/\delta\psi_m'=\delta[\vec{\Psi}_{\rm solition}']/\delta\psi_m'=i\partial_{t'}\psi_m'$,
that are explicitly expressed as,
\begin{equation}
\begin{array}{l}
\displaystyle i\partial_{t'}\psi_{\pm1}'=\left(-\frac{\partial_{x'}^2}{2}+q'+n'\pm c_s'f_z'-1\right)\psi_{\pm1}'+\frac{c_s'}{\sqrt{2}}f'_\mp\psi_0'
\\\\
\qquad\quad\displaystyle\equiv H'_{GP,\pm1}[\psi_{\pm1}',\psi_0']
\\\\
\displaystyle i\partial_{t'}\psi_0'=\left(-\frac{\partial_{x'}^2}{2}+n'-1\right)\psi_0'
+\frac{c_s'}{\sqrt{2}}\left(f'_+\psi_1+f_-'\psi_{-1}`\right)
\\\\
\qquad\quad\displaystyle\equiv H'_{GP,0}[\psi_{\pm1}',\psi_0']
\end{array}
\label{eq:GPE_dimless}
\end{equation}
where $t'=t/\tau$ is the dimensionless time. To find the soliton solution $\vec{\Psi}_{\rm soliton}'$ minimizing the soliton energy, we employ the imaginary-time propagation method to solve for the lowest state of Eq.~(\ref{eq:GPE_dimless}).  We select initial conditions of the form, 

\begin{equation}
\left(\begin{array}{l}\psi'_1
\\
\psi_0'
\\
\psi_{-1}'
\end{array}
\right)
=\left(\begin{array}{c}
A_1e^{i\theta}\sech{x'}
\\
\tanh x'
\\
A_{-1} e^{-i\theta}\sech x'
\end{array}
\right)
\end{equation}
where $A_1$ and $A_{-1}$ are complex numbers, and, owing to the rotational symmetry about $z$ axis in spin space, one can choose $\theta=0$ without loss of generality.  Furthermore, in order to carry out the imaginary-time propagation with a faster convergence, we impose the following conditions (1) $A_1=-A_{-1}\neq0$; (2) $A_1=A_{-1}\neq0$; (3) $\left|A_1\right|\neq\left|A_{-1}\right|\neq0$, and $A_1$ and $A_{-1}$ are in phase; (4) $A_1=A_{-1}=0$ such we can arrive at the designated AF-, BA-, F- or N-core more efficiently. Note that if $A_1$ and $A_{-1}$ are neither in phase nor out of phase, it turns out that the imaginary-time evolution of the initial states will end up in a soliton-free bulk solution in the P phase because such states are not spin-conserving.

In the simulations, we consider a 1D mesh where the grid size is $dx'=0.25$. The length of the mesh ranges from $L'=80$ to 720 to guarantee the boundary conditions $\psi_{\pm1}'(x=\pm L/2)=0$, $\psi_0'(x=\pm L/2)=\pm1$ or $\mp1$, as well as the Neumann boundary condition, $\partial_{x^\prime}\psi_m(x^\prime=\pm L/2)=0$ are fulfilled.
Generally speaking, $L'=80$ is sufficiently large for $q'\ge0.0375$ as the largest length scale of the core size is $\xi_v'=1/\sqrt{2q'}\sim6.3$ for $q'=0.0125$ in the flat-core approximation.
The convergence of the imaginary-time propagation is controled by the magnitude of the local error produced during the imaginary-time evolution governed by Eq.~(\ref{eq:GPE_dimless})
\begin{equation}
\delta\mu'(x')\equiv\left|\frac{(H'_{GP,m}[\psi_{\pm1}',\psi_0'])}{\psi_m'}\right|,
\end{equation}
based on the criterion max$(\delta\mu')<2.5\times10^{-13}$. 

\begin{figure}[t!]
\centering
\includegraphics[width=1\linewidth]{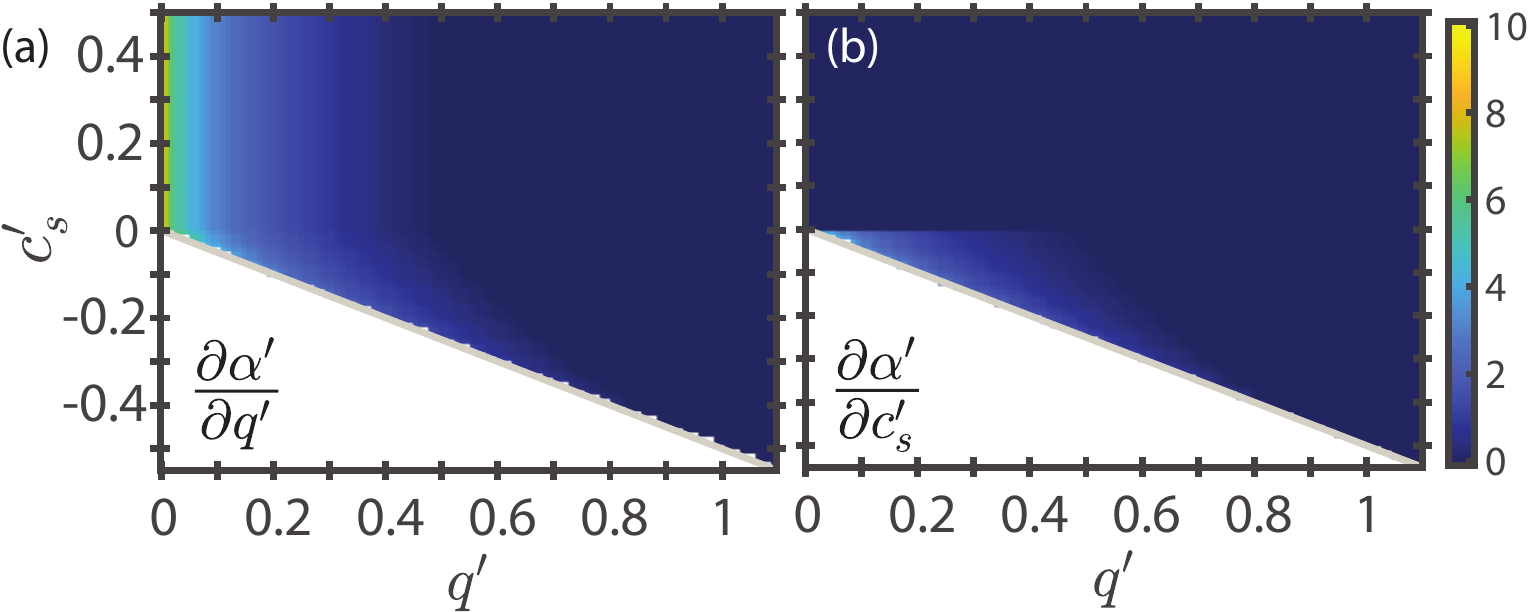}
\caption{Evaluation of the derivatives (a) $\partial\alpha^\prime/\partial q^\prime$ and (b) $\partial\alpha^\prime/\partial c_s^\prime$. }
\label{fig:dalpha}
\end{figure}

The order of phase transition between two different types of soliton cores is further examined by computing the derivatives $\partial\alpha^\prime/\partial q^\prime$ and $\partial\alpha^\prime/\partial c_s^\prime$ numerically. In Fig.~\ref{fig:dalpha}~(a), a smooth variation of $\partial\alpha^\prime/\partial q^\prime$ is clearly observed, whereas $\partial\alpha^\prime/\partial c_s^\prime$ exhibits a discontinuity at $c_s/c_d=0$ as depicted in Fig.~\ref{fig:dalpha}~(b). 
With a fixed density, the results suggest that the AF-N, F-BA and BA-N transitions are continuous or second order, and the AF-F and AF-BA transitions are discontinuous or first order.
\section{: P\"oschl-Teller theory}\label{sec:effective-N-core}
The time-independent Schr\"odinger equation for the  P\"oschl-Teller potential is given by~\cite{Diaz1999,PTequation}
\begin{equation}
\epsilon_n\phi_n(x)=\left[-\frac{\hbar^2}{2M}\dfrac{d^2}{dx^2}-\frac{g\alpha}{4}\sech^2(\alpha x)\right]\phi_n(x),
\label{eq:appendix_PT_equation}
\end{equation}
where $\epsilon_n$ and $\phi_n$ denote the $n$-th energy eigenvalue and eigenstate. 
Given that $\alpha>0$, $g>0$ and $\lambda=(1/2)(1+\sqrt{1+4Mg/\hbar^2\alpha})>1$, there exists bound-state solutions associated with the energy eigenvalue 
\begin{equation}
\epsilon_n=-\frac{\hbar^2\alpha^2}{4M}\left(
\lambda-1-n\right)^2,
\label{eq:eigen_E}
\end{equation}
where $n\in\mathbb{N}$ and $0\le n<\lambda-1$. The details of solving the eigenvalue equation Eq.~(\ref{eq:appendix_PT_equation}) can be found in Ref.~\cite{Diaz1999} and problem 5 in Sec. 23 in Ref.~\cite{Landau1980}.
 
Replacing the P\"oschl-Teller potential in Eq.~(\ref{eq:appendix_PT_equation}) with the effective potentials $V_\pm(x)=-U_\pm\sech^2(x/\xi)$, we have     
\begin{equation}
\displaystyle\alpha\rightarrow\frac{1}{|\xi|}>0
\quad\textrm{,}
\quad\displaystyle g\rightarrow2|\xi|U_\pm,
\end{equation}
such that the corresponding $\lambda$ parameters are given by
\begin{equation}\begin{array}{rl}
\lambda_\pm=&\displaystyle\frac{1}{2}\left(
1+\sqrt{1+\frac{8M\xi^2}{\hbar^2}U_\pm}
\right)
\\\\
=&\displaystyle\frac{1}{2}\left(
1+\sqrt{9+8(1\pm1)\frac{c_s}{c_d}}
\right).
\end{array}
\end{equation}

Consequently, the bound-state solutions exist only when the following conditions are met: $c_s/c_d>-1/2$ for $g>0$, and $(1\pm1)({c_s}/{c_d})>-8$ for $\lambda_\pm>1$. Denote $\epsilon_{\pm, n}$ as the $n$-th energy eigenvalues for the potential $V_\pm(x)$ respectively. For $V_-$, it follows that $\lambda_-=2>1$, and according to Eq.~(\ref{eq:eigen_E}), there is only one bound-state $\Phi_-$ with energy level $\epsilon_{-,0}=-{\mu}/{2}$ exists. Thus Eq.~(\ref{eq:linearPhi}) has non-trivial solution if and only if $E_-=\epsilon_{-,0}$, implying that 
\begin{equation}
\frac{q}{\mu}=\frac{1}{2}.
\label{eq:appendixe_e_-}
\end{equation}
Next, we solve the eigenvalue equation of $\Phi_+$ in the regime, $-q/2\mu<c_s/c_d<0$, where the BA-core is favorable. This yields $1<\lambda_+<2$, implying that only one bound-state solution exists, which has the energy level 
\begin{equation}
\epsilon_{+,0}=-\frac{\mu}{8}\left[\sqrt{9+16\frac{c_s}{c_d}}-1\right]^2
\end{equation}
Likewise, Eq.~(\ref{eq:linearPhi}) has non-trivial solution if and only if $E_+=\epsilon_{+,0}$, which then gives the equation of the BA-N phase boundary  
\begin{equation}
\dfrac{c_{s}}{c_{d}} =1-\dfrac{5}{2}\dfrac{q}{\mu } +\dfrac{q^{2}}{\mu ^{2}}.
\label{eq:appendixe_e_+}
\end{equation}
Equations~(\ref{eq:appendixe_e_-}) and (\ref{eq:appendixe_e_+}) represent the boundaries of the AF-N and BA-N phase transitions respectively, which are mutually exclusive. This rules out the possibility of the F-N transition as none of $\Phi_\pm$ vanish in the F-core phase. Since Eq.~(\ref{eq:appendixe_e_+}) are well-founded in the regime $-q/2\mu<c_s/c_d<0$ and intersects with Eq.~(\ref{eq:appendixe_e_-}) at $(q/\mu,c_s/c_d)=(1/2,0)$, it follows that the whole border separating the N-core phase from other possible phases can be mapped out by piecewisely joining the lower part ($c_s<0$) of Eq.~(\ref{eq:appendixe_e_+}) 
to the upper part ($c_s>0$) of Eq.~(\ref{eq:appendixe_e_-}) at the point $(q/\mu,c_s/c_d)=(1/2,0)$ , as shown in Fig.~\ref{fig:1}.


\section{: The Ginzburg-Landau-like approach}
\label{sec:critical_behaviour}

\begin{figure}[t!]
\centering
\includegraphics[width=1\linewidth]{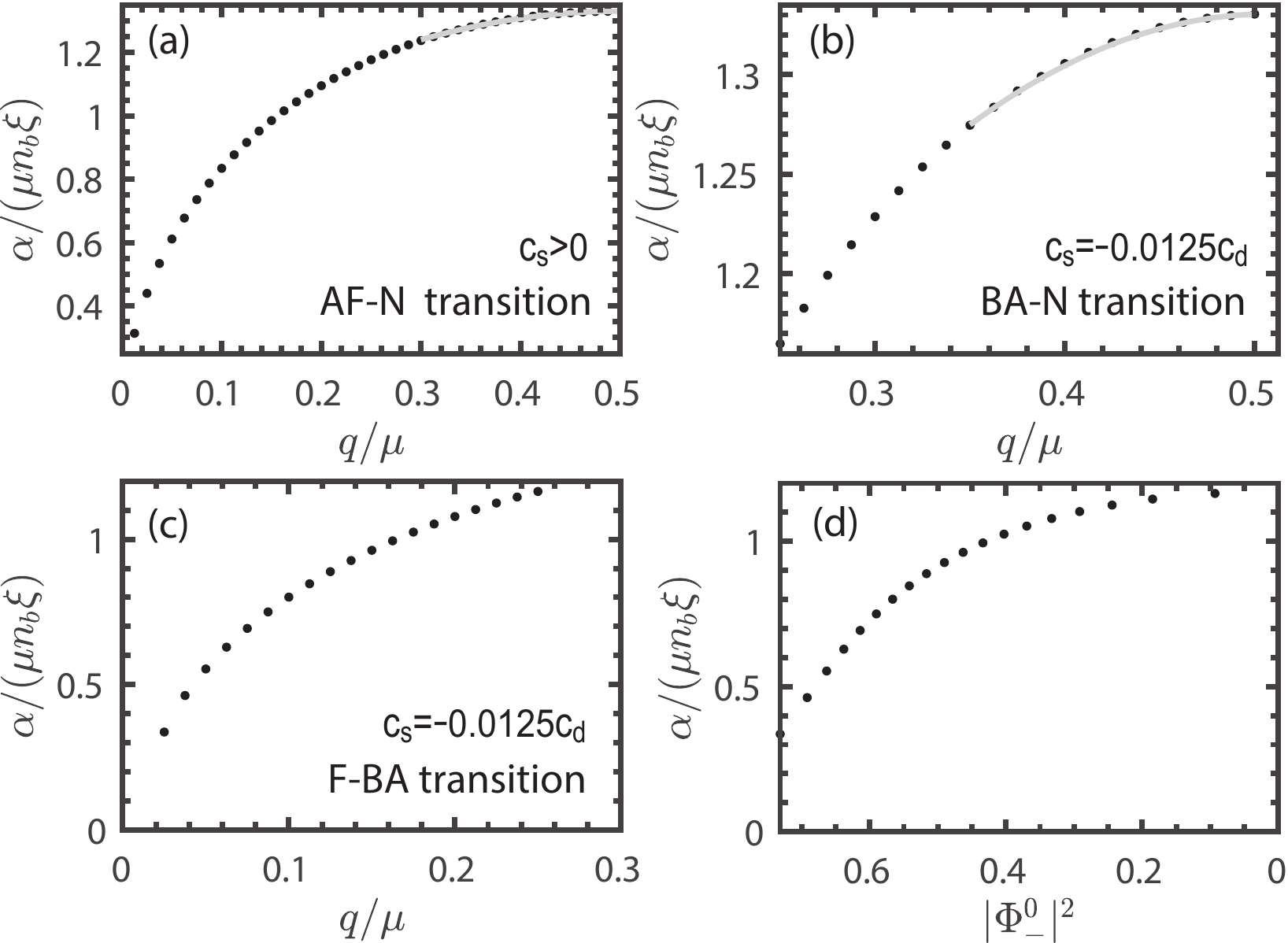}
\caption{
The $\alpha_c$ (black dots) as a function of $q/\mu$ is shown for (a) AF-N, (b) BA-N and (c) F-AB transitions. The grey solid lines in (a) and (b) indicate the curve fitting using Eq.~(\ref{eq:GL_STC}). In (d), the $\alpha_c$ as a function of $\left|\Phi_-^0\right|^2$ for the F-BA transition in (c) is illustrated, which appears as a parabolic-like curve. 
}
\label{fig_app:alpha}
\end{figure}

In the framework of mean-field theory, the second-order phase transition, which is characterized by the vanishing of order parameter at the critical point, can be described by the Ginzburg-Landau~(GL) theory.  In view of the continuous nature of the AF-N,F-BA and BA-N phase transitions, by analogy with the GL theory, the soliton energy is expanded in terms of the order parameter in the critical region for $q\leq q_c$ by
\begin{equation}
\alpha=\alpha_0-\alpha_1(q-q_c)\left|\varphi\right|^2-\alpha_2\left|\varphi\right|^4
\label{eq:GL_STC_0}
\end{equation}
where $\varphi$ is the order parameter, and specifically, $\varphi=\Phi_+^0$ for the BA-N and $\varphi=\Phi_-^0$ for the AF-N and F-BA transitions. The coefficients $\alpha_i$ $(i=0,1, 2)$ in Eq.~(\ref{eq:GL_STC_0}) are all positive and determined by $\Phi_\pm$ and $\Phi_0$. For the BA-N and  AF-N transitions, $\alpha_0$ can be exactly determined and is explicitly given by $\alpha_0=4\mu n_b\xi/3$. Minimization of $\alpha$ with respect to $\varphi$ yields
\begin{equation}
\left|\varphi\right|^2=\frac{\alpha_1}{\alpha_2}\left|q-q_c\right|=\beta\left|q-q_c\right|,\textrm{  for  } q\leq q_c,
\label{eq:GL_order}
\end{equation}
with $\beta=\alpha_1/\alpha_2$. Substituting Eq.~(\ref{eq:GL_order}) into Eq.~(\ref{eq:GL_STC_0}), we get 
\begin{equation}
\alpha=\alpha_0-2\alpha_{1}\beta(q-q_c)^2,\textrm{   for  } q\leq q_c.
\label{eq:GL_STC}
\end{equation}
In Fig.~\ref{fig_app:alpha}~(a) and (b), the calculated $\alpha_c$ are well fitted by the quadratic function of Eq.~(\ref{eq:GL_STC}) for AF-N and BA-N transitions, with $\alpha_1=0.56$ and $0.57$ respectively.

\subsection{AF-N transition} In this case, $\varphi=\Phi_-^0$, and the two parameters $\beta=2n_b/\mu$ and $q_c=\mu/2$ are exactly determined based on the results obtained by P\"oschl-Teller approach. Since $\left|\varphi\right|^2=\left|\Phi_-^0\right|^2=n(x=0)=n_0$, it follows that $n_0\propto\left|q-q_c\right|$, in other words, the central density linearly decreases to zero in the critical region for $q\leq q_c$, which is clearly illustrated in Fig.~\ref{fig:3}~(a).

\subsection{BA-N transition} In this case, $\varphi=\Phi_+^0$. However, as $\alpha_i$ become $c_s$-dependent due to the spin-spin interaction, $\beta$ and $q_c$ can only be determined numerically. For example, given $c_s=-0.0125c_d$, we find $\beta=1.97n_b/\mu$ and $q_c=0.51\mu$. Like the AF-N phase transition,  the central density exhibits linear decline near the critical point, $n_0=\left|\varphi\right|^2=\left|\Phi_+^0\right|^2\propto\left|q-q_c\right|$ for $q\leq q_c$ as shown in Fig.~\ref{fig:3}~(b). In what follows, we shall show that, in addition to the central density, the maximal spin density also exhibit power law behavior near the critical point. Recall that for the BA-core, $f_z(x)=0$ and $f_\perp(x)=2\Phi_0\Phi_+$. Defining $ \text{max}\left[f_{\perp}(x)\right]=f_{\perp}^\text{max}$ and from Fig.~\ref{fig:2}, we see that $f_{\perp}^\text{max}$ always occurs around the core. Assuming that $f_{\perp}^\text{max}$ occurs at $x=x_f$ in the proximity of the core center $x=0$, and expanding $f_{\perp}^\text{max}$ to the first order, we obtain
\begin{equation}
f_{\perp}^\text{max} \approx2x_f\left(\frac{d\Phi_0}{dx}\Phi_+ \right)_{x=0}\propto\Phi_+^0\propto\left|q-q_c\right|^{1/2}.
\label{eq:app_fperp}
\end{equation}
The above behavior is numerically verified and is shown in Fig.~\ref{fig:3}~(b) in good agreement.

\subsection {F-BA transition} In this case, $\varphi=\Phi_-^0$, and the two parameters $\beta_1$ and $q_c$ are to be determined numerically. As shown in the inset of Fig.~\ref{fig:3}~(c), in the F-BA transition, $|\Phi_-^0|^2$ declines linearly in the critical region and becomes zero at $q_c$ as expected, whereas $\Phi_+^0$ linearly decreases to a nonzero value at $q_c$. Thus, in addition to the decline of the order parameter, $\left|\Phi_-^0\right|^2=\beta\left|q-q_c\right|$, we assume $\left|\Phi_+^0\right|^2=B_0+B_1\left|q-q_c\right|$, with $B_0>0$ the total central density at $q_c$, and $B_1>0$ the descent rate. Thus the total central density is given
\begin{equation}
n_0=\left|\Phi_+^0\right|^2+\left|\Phi_-^0\right|^2=B_0+(\beta_1+B_1)\left|q-q_c\right|,
\end{equation}
The coefficients $\beta$, $B_0$ and $B_1$ are determined by the linear fit to $n_0$. Given $c_s=-0.0125c_d$, we have $q_c=0.25\mu$, $\beta= 1.04n_b/\mu$, $B_0=0.5$ and $B_1=0.99$. 
Likewise, we define $\text{max}\left[f_z(x)\right]=f_z^\text{max}$, and from Fig.~\ref{fig:2}, it follows that $f_z^\text{max}=f_z(x=0)$. Consequently, in the critical region for $q\leq q_c$, we get
\begin{equation}
f_{z}^\text{max}=2\Phi_+^{0}\Phi_-^0\propto\sqrt{\beta\left|q-q_c\right|\left(B_0+B_1\left|q-q_c\right|\right)},
\label{eq:app_fz}
\end{equation}
and from Eq.~(\ref{eq:app_fperp}), it follows that
\begin{equation}
f_{\perp}^\text{max}\propto\Phi_+^0\propto\sqrt{B_0+B_1\left|q-q_c\right|}.
\end{equation}
\end{appendices}
\bibliographystyle{unsrt}
\bibliographystyle{prsty}

\begin{thebibliography}{10}

\bibitem{2000Bunkov}
Y.~M. Bunkov and H. Godfrin, {\em Topological defects and the non-equilibrium
  dynamics of symmetry breaking phase transitions} (Springer Science \&
  Business Media, ADDRESS, 2012), Vol.~549.

\bibitem{Kibble76}
T. Kibble, Journal of Physics A: Mathematical and General {\bf 9},  1387
  (1976).

\bibitem{Zurek96}
W. Zurek, Physics Reports {\bf 276},  177   (1996).

\bibitem{2013Lamporesi}
G. Lamporesi {\it et~al.}, Nature Physics {\bf 9},  656  (2013).

\bibitem{Liu2018}
I.-K. Liu {\it et~al.}, Communications Physics {\bf 1},  24  (2018).

\bibitem{2010Damski}
B. Damski and W.~H. Zurek, Physical review letters {\bf 104},  160404  (2010).

\bibitem{2003Volovik}
G.~E. Volovik, {\em The universe in a helium droplet} (Oxford University Press
  on Demand, ADDRESS, 2003), Vol.~117.

\bibitem{2013Vollhardt}
D. Vollhardt and P. W{\"o}lfle, {\em The superfluid phases of helium 3}
  (Courier Corporation, ADDRESS, 2013).

\bibitem{Volovik2020}
G.~E. Volovik and K. Zhang, Phys. Rev. Research {\bf 2},  023263  (2020).

\bibitem{Dalfovo1999}
F. Dalfovo, S. Giorgini, L.~P. Pitaevskii, and S. Stringari, Rev. Mod. Phys.
  {\bf 71},  463  (1999).

\bibitem{Pethick2008}
C.~J. Pethick and H. Smith, {\em Bose-Einstein Condensation in Dilute Gases}
  (Cambridge University Press, ADDRESS, 2008).

\bibitem{Kawaguchi2012}
Y. Kawaguchi and M. Ueda, Physics Reports {\bf 520},  253  (2012).

\bibitem{2003Leahrdt}
A. Leanhardt {\it et~al.}, Physical review letters {\bf 90},  140403  (2003).

\bibitem{1995Parts}
{\"U}. Parts {\it et~al.}, Physical review letters {\bf 75},  3320  (1995).

\bibitem{Kang2018}
S. Kang, S.~W. Seo, H. Takeuchi, and Y. Shin, Phys. Rev. Lett. {\bf 122},
  095301  (2019).

\bibitem{AFandP}
The antiferromagnetic and polar phases are also called easy-plane polar (EPP)
  and easy-axis polar (EAP) phases, respectively.

\bibitem{Anderson2001}
B.~P. Anderson {\it et~al.}, Phys. Rev. Lett. {\bf 86},  2926  (2001).

\bibitem{Burger1999}
S. Burger {\it et~al.}, Physical Review Letters {\bf 83},  5198  (1999).

\bibitem{Busch2001}
T. Busch and J.~R. Anglin, Phys. Rev. Lett. {\bf 87},  010401  (2001).

\bibitem{Frantzeskakis2010}
D. Frantzeskakis, Journal of Physics A: Mathematical and Theoretical {\bf 43},
  213001  (2010).

\bibitem{2011Witkowska}
E. Witkowska, P. Deuar, M. Gajda, and K. Rz\k{a}\ifmmode~\dot{z}\else
  \.{z}\fi{}ewski, Phys. Rev. Lett. {\bf 106},  135301  (2011).

\bibitem{2005Li}
L. Li {\it et~al.}, Physical Review A {\bf 72},  033611  (2005).

\bibitem{2006Uchiyama}
M. Uchiyama, J. Ieda, and M. Wadati, Journal of the Physical Society of Japan
  {\bf 75},  064002  (2006).

\bibitem{2008Nistazakis}
H. Nistazakis {\it et~al.}, Physical Review A {\bf 77},  033612  (2008).

\bibitem{2018Bersano}
T. Bersano {\it et~al.}, Physical review letters {\bf 120},  063202  (2018).

\bibitem{Meng2019}
Y.-H.~Q. Ling-Zheng Meng~and and L.-C. Zhao, arXiv:1912.00182.

\bibitem{Becker2008}
C. Becker {\it et~al.}, Nature Physics {\bf 4},  496  (2008).

\bibitem{Weller2008}
A. Weller {\it et~al.}, Phys. Rev. Lett. {\bf 101},  130401  (2008).

\bibitem{Theocharis2010}
G. Theocharis {\it et~al.}, Phys. Rev. A {\bf 81},  063604  (2010).

\bibitem{Kevrekidis2016}
P. Kevrekidis and D. Frantzeskakis, Reviews in Physics {\bf 1},  140  (2016).

\bibitem{Farolfi2019}
A. Farolfi {\it et~al.},  Phys. Rev. Lett. {\bf{125}}, 030401 (2020).

\bibitem{Chai2019}
X. Chai {\it et~al.}, Phys. Rev. Lett. {\bf{125}}, 030402 (2020).

\bibitem{Lieb2000}
E.~H. Lieb, R. Seiringer, and J. Yngvason, Phys. Rev. A {\bf 61},  043602
  (2000).

\bibitem{mean_field}
{T}he accuracy and limitations of GP mean-field treatment is also addressed in
  Ref.~\cite{ Dalfovo1999}. As is pointed out, with a temperature $T$ up to
  $0.4T_c$, i.e., $T\lessapprox0.4T_c$, where $T_c$ is the BEC transition
  temperature, the GPE describes reasonably well the static and dynamic
  properties of a weakly interacting Bose gas, and the discrepancy between the
  GPE modeling and the experimental observations is typically less than a few
  percent.

\bibitem{stationaryHere}
In this work, we neglect the soliton solutions with local current density. Such
  a soliton solution has been obtained in Rabi-coupled two-component BECs; A.
  Usui and H. Takeuchi, Phys. Rev. A {\bf 91} 063635 (2015).

\bibitem{Salasnich2006}
L. Salasnich and B.~A. Malomed, Phys. Rev. A {\bf 74},  053610  (2006).

\bibitem{Landau1980}
L.~D. Landau and E. Lifshitz, {\em Statistical Physics, Third Edition, Part 1:
  Volume 5 (Course of Theoretical Physics, Volume 5)} (Butterworth-Heinemann,
  ADDRESS, 1980).

\bibitem{NumericalDetails1}
The details of numerical computations and the derivative of soliton energy with respect to $q/\mu$ and $c_s/c_d$ are included in
  Sec.~\ref{sec:numerical_process} of the supplemental materials.


\bibitem{Hayashi2013}
S. Hayashi, M. Tsubota, and H. Takeuchi, Phys. Rev. A {\bf 87},  063628
  (2013).

\bibitem{Diaz1999}
J. Diaz, J. Negro, L. Nieto, and O. Rosas-Ortiz, Journal of Physics A:
  Mathematical and General {\bf 32},  8447  (1999).

\bibitem{PTequation}
See, e.g., Problem 5 in Sec. 23 of Ref.~\cite{Landau1980}.

\bibitem{PTtheory}
Some properties of the bound-state solutions are included in
  Sec.~\ref{sec:effective-N-core} of the appendix.

\bibitem{flat_core_var_para}
The opparations of $\xi_v\to -\xi_v$ and $\varphi_v\to \varphi_v+\pi$ cause a
  flip of magnetazation as $(f_\bot\to -f_\bot)$, which does not change the
  soliton tension coefficient.

\bibitem{Prfer2018}
M. Pr\"{u}fer {\it et~al.}, Nature {\bf 563},  217  (2018).

\bibitem{JimnezGarca2019}
K. Jim{\'{e}}nez-Garc{\'{\i}}a {\it et~al.}, Nature Communications {\bf 10},
  (2019).

\bibitem{Huh2020}
{S}. Huh {\it{et al.}}, arXiv:2006.06228.

\bibitem{Lannig2020}
{S}. Lannig, {\it{et. al}}, arXiv:2005.13278

\end{thebibliography}

\end{document}